\documentstyle[12pt,psfig]{article}

\newcommand{\m}{\medbreak}
\newcommand{\no}{\noindent}
\newcommand{\EQ}{\begin{equation}}
\newcommand{\eq}{\end{equation}}
\newcommand{\EQA}{\begin{eqnarray}}
\newcommand{\eqa}{\end{eqnarray}}

\def\subfigureA#1{
\leavevmode
\hbox{#1}
}
\def\pr#1#2#3{ Phys. Rev. {\bf{#1}} (#2) #3}
\def\prl#1#2#3{ Phys. Rev. Lett. {\bf{#1}} (#2) #3}
\def\pl#1#2#3{ Phys. Lett. {\bf{#1}} (#2) #3 }

\def\np#1#2#3{ Nucl. Phys. {\bf{#1}} (#2) #3}
\def\zp#1#2#3{ Z. Phys. {\bf{#1}} (#2) #3}

\begin{document}

\begin{titlepage}
\vspace{0.2in}
\vspace*{1.5cm}
\begin{center}
{\large \bf Polarized beams at HERA : \\ 
analyzing the chiral structure
of contact interactions
\\} 
\vspace*{1.2cm}
{\bf J.-M. Virey}{$^1$}  \\ \vspace*{1cm}
Centre de Physique Th\'eorique$^{\ast}$, C.N.R.S. - Luminy,
Case 907\\
F-13288 Marseille Cedex 9, France\\ \vspace*{0.2cm}
and \\ \vspace*{0.2cm}
Universit\'e de Provence, Marseille, France\\
\vspace*{1.8cm}
{\bf Abstract \\}
\end{center}
In the context of HERA with polarized lepton and proton beams, we
explore the sensitivity of the collider to contact interactions.
We emphasize that the measurement of longitudinal
spin asymmetries in such a polarized context could give some crucial informations
on the chiral structure of these hypothetical new interactions.
\vspace{0.8cm}

\no {\it To appear in the proceedings of the "2nd Zeuthen Spin Physics Workshop: Theory 
meets Experiment", Zeuthen,
Germany, September 1997; Working group on "Physics with Polarized Protons at HERA",
CERN, October 1997. \\
}

\vfill
\begin{flushleft}
\m\no
Number of figures : 4\\

\m\no
October 1997\\
CPT-97/P.3542\\
\m\no
anonymous ftp or gopher : cpt.univ-mrs.fr

------------------------------------\\
$^{\ast}$Unit\'e Propre de Recherche 7061

{$^1$} E-mail : Virey@cpt.univ-mrs.fr
\end{flushleft}
\end{titlepage}


\section{Introduction}

\vspace{1mm}
\noindent
A contact interaction (CI) between electron and quark could mimic any
new physics manifestation in $eq \rightarrow eq$ scattering \cite{Bargerct2}.
It can represent a common substructure between electron and quark
\cite{ELP83}, or the
exchange of a $Z'$ \cite{Bargerct1,pcjmv} or of a leptoquark \cite{Martyn91,Bargerct2}, 
if the boson mass is such that $M \gg \sqrt{s}$.
We consider a new $eq$ CI,
which is normalized to a certain
energy scale $\Lambda$, with the following effective 
Lagrangian \cite{ELP83,Ruckl84}:
\EQA\label{Leq}
{\cal L}_{eq} = \sum_q ( \eta^q_{LL}(\bar e_L \gamma_{\mu} e_L) 
(\bar q_L \gamma^{\mu} q_L) +  \eta^q_{RR}(\bar e_R \gamma_{\mu} e_R) 
(\bar q_R \gamma^{\mu} q_R)  \nonumber 
  \\   + \eta^q_{LR}(\bar e_L
\gamma_{\mu} e_L)  (\bar q_R \gamma^{\mu} q_R) + \eta^q_{RL}(\bar e_R
\gamma_{\mu} e_R)  (\bar q_L \gamma^{\mu} q_L) )  
\eqa

\no with $\eta^q_{ij}= \epsilon g^2/(\Lambda^q_{ij})^2$ where
$g^2=4 \pi$ and $\epsilon=\pm 1$. The sign $\epsilon$ caracterises the 
nature of the interferences with the Standard Model (SM) amplitudes. 
The four subscripts $LL$, $RR$, $LR$ and $RL$ caracterise the chiral
structure of the new interaction. Assuming the existence of 
contact terms for the first generation
of quarks and $u$-$d$ "universality" {\it i.e.} $\eta_{ij}=\eta^u_{ij}=\eta^d_{ij}$,
these four chiralities, along with the sign $\epsilon$, define eight individual
models. The CI could correspond to one of these models or
to any combination of them.

These individual models are constrained by several experiments involving
$eq$ scattering (see \cite{Bargerct2} for a nice review). In particular, the
atomic parity violation experiments on Cesium atoms  give some bounds of the order
of $\Lambda \sim 10$ TeV \cite{Bargerct2,Deandrea}. However it appears that it is easy to find
some combinations of the chiralities which evade these constraints 
\cite{Nels}. Nevertheless, for simplicity, we will consider the eight models
individually and then observe the effects of more complicated models at the end.

The H1 and ZEUS collaborations at HERA have observed an excess of
events, in comparison with the SM expectations, at high $Q^2$, in the deep
inelastic positron-proton cross section $\sigma_+ \equiv d\sigma / dQ^2
(e^+p \rightarrow e^+ X)$ \cite{HERAe}. This excess could be interpreted as a manifestation
of new physics (see \cite{Alt1} for a very recent review) : leptoquarks, squarks with
R parity violation or contact interactions. We will give a remark on leptoquarks
in the conclusion, but now we concentrate on CI. The HERA anomaly could be
interpreted as a CI with a scale $\Lambda \sim 3-4\, TeV$, in the
up quark sector. Note that, since the lepton beam is made of positrons, the
cross section $\sigma_+$ is sensitive to the chiralities $LR^\pm$ and/or $RL^\pm$
where $\pm$ correspond to $\epsilon$.

With the present values for the parameters
of the HERA experiments \cite{HERAe}, but with higher integrated luminosities
($L_{e^-}=L_{e^+}=1\, fb^{-1}$), we can show \cite{jmvp,pcjmv} that 
the cross section measurements
can probe an energy scale of the order of $7\, TeV$ for constructive
interferences ($\epsilon=+1$), and of the order of $6\, TeV$ in the destructive case. 
We conclude that the present HERA anomaly will be soon confirmed or
invalidated. However it appears that electron {\it and} positron beams
are necessary to cover all the possible chiralities of the new interaction.
The comparison of the two cross section $\sigma_-$ and $\sigma_+$
allow the distinction of two classes of chiralities : $(LL^\pm,RR^\pm)$
and $(LR^\pm,RL^\pm)$ \cite{martyn,Bargerct1,jmvp}. But we have to note that
cross section measurements are unable to discriminate between chiralities
within each class.

Now, we want to emphasize that the measurement of some spin asymmetries,
defined in the context of HERA with polarized lepton beams and also with a
polarized proton beam, could give some very important information
on the chiral structure of the new interaction. Note that lepton
polarization is part of the HERA program, and that proton polarization is the
aim of this workshop !\\
\no The evaluation of the cross sections, of the asymmetries and the
corresponding errors
is made with the following parameters :
$\sqrt{s} = 300\, GeV$, $0.01 < y < 0.95$, 
$L_{e^\pm} = 250\, pb^{-1} $ per spin configuration.
This choice for the integrated luminosity is maybe too high, but if we divide this value 
by a factor two, the bounds given in the following 
decrease by $\sim$15\%, which is in remarkable agreement with the scaling law given in
\cite{Leike}. Concerning the $Q^2$ resolution we take 
$\Delta Q^2/Q^2
= 34.3\,\%$ and $Q^2_{min} = 200\, GeV^2$. We note that the
GRSV polarized parton distributions \cite{GRSV} are used for the calculations. 
This choice corresponds to a conservative attitude since for this set 
of distributions the quarks are
weakly polarized in comparison with other sets which are currently used, like GS96 or BS
\cite{param}. As a consequence, the spin effects are weaker giving smaller bounds on 
$\Lambda$. Note that this uncertainty will be strongly reduced thanks to the spin
asymmetries measurements at the RHIC-BNL polarized $pp$ collider, 
for $\gamma$, jets and $W^{\pm}$ productions \cite{jsjmv}.
The degrees of polarization of the beams are taken such that
$P_{e^-}=P_{e^+}=P_{p}=70 \, \%$. 
Finally, we have chosen a total systematical error of 10\% for the asymmetries :
$\Delta A_{syst}/A = 10\, \%$.

\section{Results}

\vspace{1mm}
\noindent
We have simulated sixty spin asymmetries 
that we can construct with the eight independent cross sections :
\EQ
\sigma_-^{--} \;\;\;\;\sigma_-^{++} \;\;\;\;\sigma_-^{-+} \;\;\;\;\sigma_-^{+-} 
\;\;\;\;\; and \;\;\;\;\;
\sigma_+^{--} \;\;\;\;\sigma_+^{++} \;\;\;\;\sigma_+^{-+} \;\;\;\;\sigma_+^{+-}
\eq
\no where $\sigma_t^{\lambda_e \lambda_p} \equiv (d\sigma_t/dQ^2)^{\lambda_e 
\lambda_p}$, where $t$ refers to the electric charge of the colliding lepton and
$\lambda_e, \lambda_p$ are the helicities of the lepton and the proton, respectively.

\no It appears that the observables which are the most sensitive to the presence 
of the CI, are the Parity Violating (PV) spin asymmetries :
\EQ\label{defALLPV}
A_{LL}^{PV}(e^-)\; =\; \frac{\sigma^{--}_- \, -\, \sigma^{++}_-}{\sigma^{--}_- \, 
+\, \sigma^{++}_-}
\;\;\;\;\;\; and \;\;\;\;\;\;
A_{LL}^{PV}(e^+)\; =\; \frac{\sigma^{--}_+ \, -\, \sigma^{++}_+}{\sigma^{--}_+ \, 
+\, \sigma^{++}_+} \;\; ,
\eq
\no and the "mixed" charge-spin asymmetry :
\EQ\label{defB22}
B_2^2\; =\; \frac{\sigma_-^{++} \, -\, \sigma_+^{++}}{\sigma_-^{++} \, +\, 
\sigma_+^{++}} \;\; .
\eq

\no The Parity Conserving (PC) spin asymmetries, which are defined when only the
proton spin is flipped, are also relevant, in particular for
the chiral structure analysis (see below) :
\EQ
A_2^1\; =\; \frac{\sigma^{--}_- \, -\, \sigma^{-+}_-}{\sigma^{--}_- \, 
+\, \sigma^{-+}_-}
\;\;\;\;\;\;\;\;\;\;\;\;\; and \;\;\;\;\;\;\;\;\;\;\;\;\;
A_2^2\; =\; \frac{\sigma^{++}_- \, -\, \sigma^{+-}_-}{\sigma^{++}_- \, 
+\, \sigma^{+-}_-}\;\; ,
\eq
\EQ
A_2^3\; =\; \frac{\sigma^{--}_+ \, -\, \sigma^{-+}_+}{\sigma^{--}_+ \, 
+\, \sigma^{-+}_+}
\;\;\;\;\;\;\;\;\;\;\;\;\; and \;\;\;\;\;\;\;\;\;\;\;\;\;
A_2^4\; =\; \frac{\sigma^{++}_+ \, -\, \sigma^{+-}_+}{\sigma^{++}_+ \, 
+\, \sigma^{+-}_+} \;\; .
\eq

\no Using a $\chi^2$ analysis we obtain the bounds presented in Table 1.

\vspace{0.5cm}

\begin{center}
\begin{tabular}{|c||c|c|c|c||c|c|c|c|}
\hline
$\Lambda$ (TeV) & $\Lambda_{LL}^+$&$\Lambda_{RR}^+$&$\Lambda_{LR}^+$
&$\Lambda_{RL}^+$&$\Lambda_{LL}^-$&$\Lambda_{RR}^-$&$\Lambda_{LR}^-$&$\Lambda_{RL}^-$\\
\hline
$A_{LL}^{PV}(e^\pm)\, or \, B_2^2$& $6.6$ & $7.2$ 
& $7.0$ & $7.0$ & $6.3$ & $7.0$ & $6.8$ & $6.7$ \\
\hline
$A_{2}^{PC,x}$& $5.3^1$ & $5.2^2$ & $5.6^4$ & $5.2^3$ 
& $5.6^1$ & $5.5^2$ & $5.8^4$ & $5.5^3$ \\
\hline
\end{tabular} 
\end{center}
\begin{center}
Table 1: Limits on $\Lambda$ at 95\% CL.
\end{center}

\no We find that the limits are comparable to the unpolarized case \cite{jmvp,martyn}.
They are slightly better for destructive interferences 
($\epsilon=-1$). The sensitivity of the PC asymmetries is $\sim$20\% smaller than the PV one's.

\begin{figure}
\vspace*{-2.cm}
\begin{tabular}[t]{c c}
\vspace*{-1.5cm}
\centerline{\subfigureA{\psfig{file={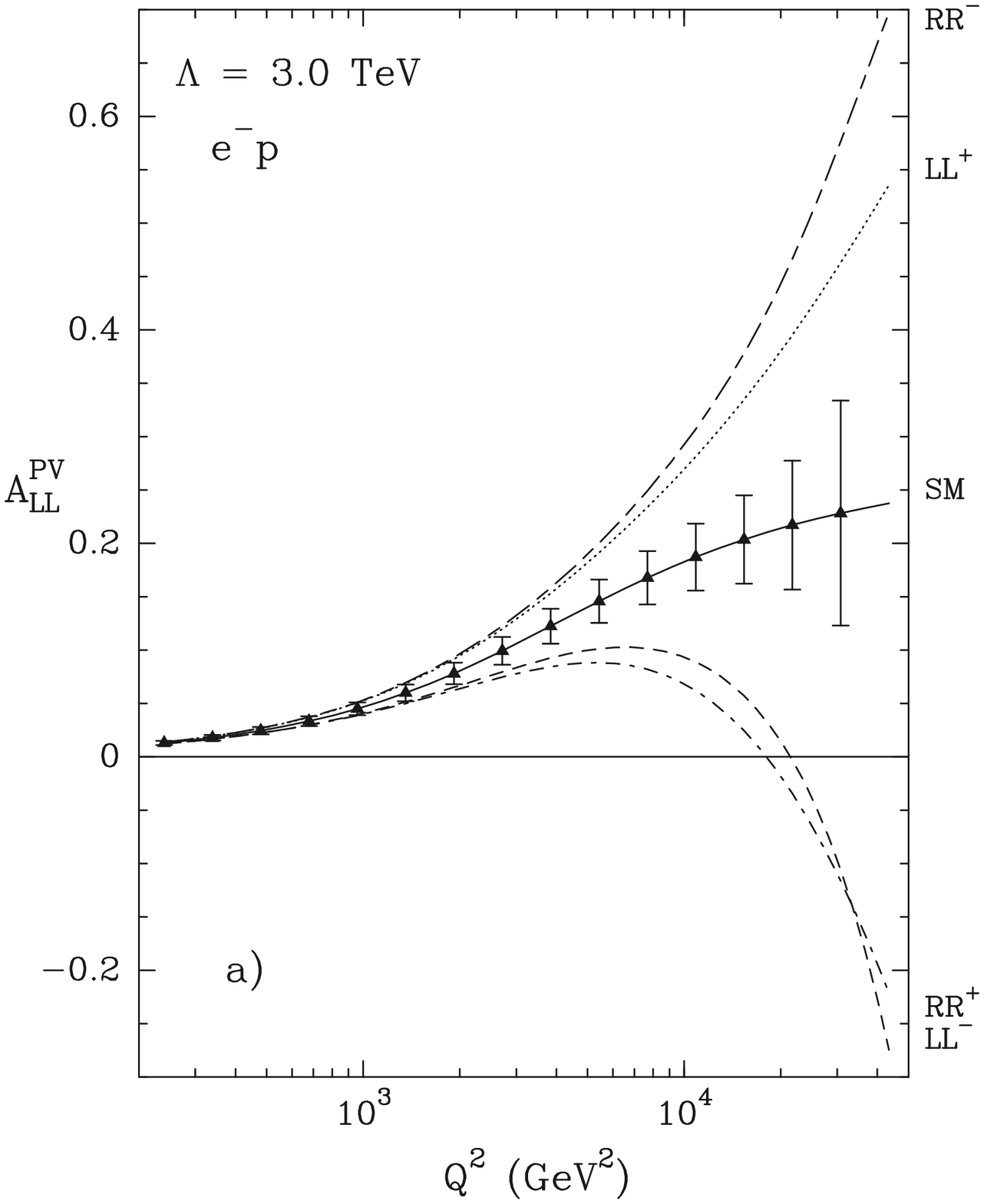},width=9truecm,height=13truecm}}\subfigureA{\psfig{file={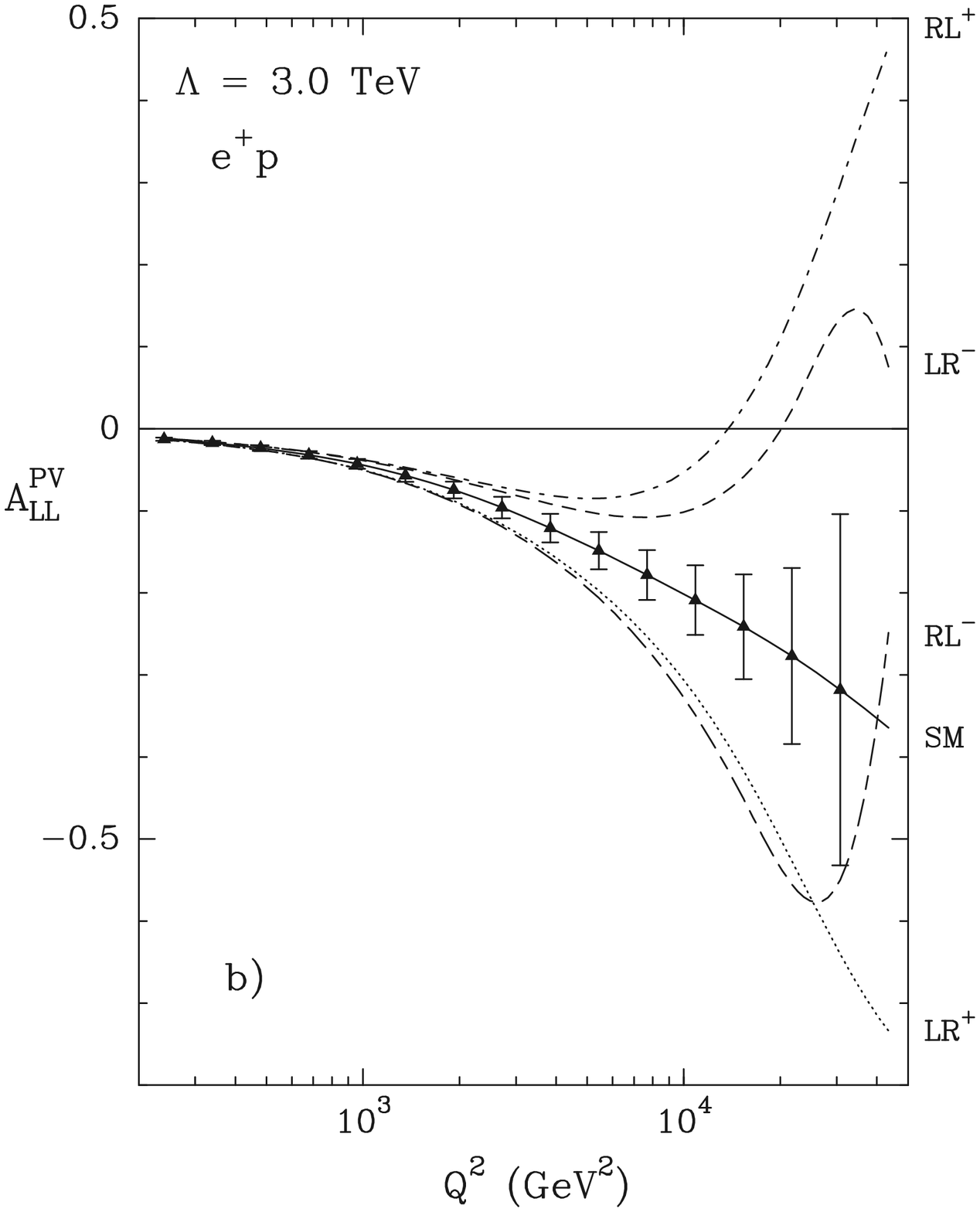},width=9truecm,height=13truecm}}}\\
\centerline{\subfigureA{\psfig{file={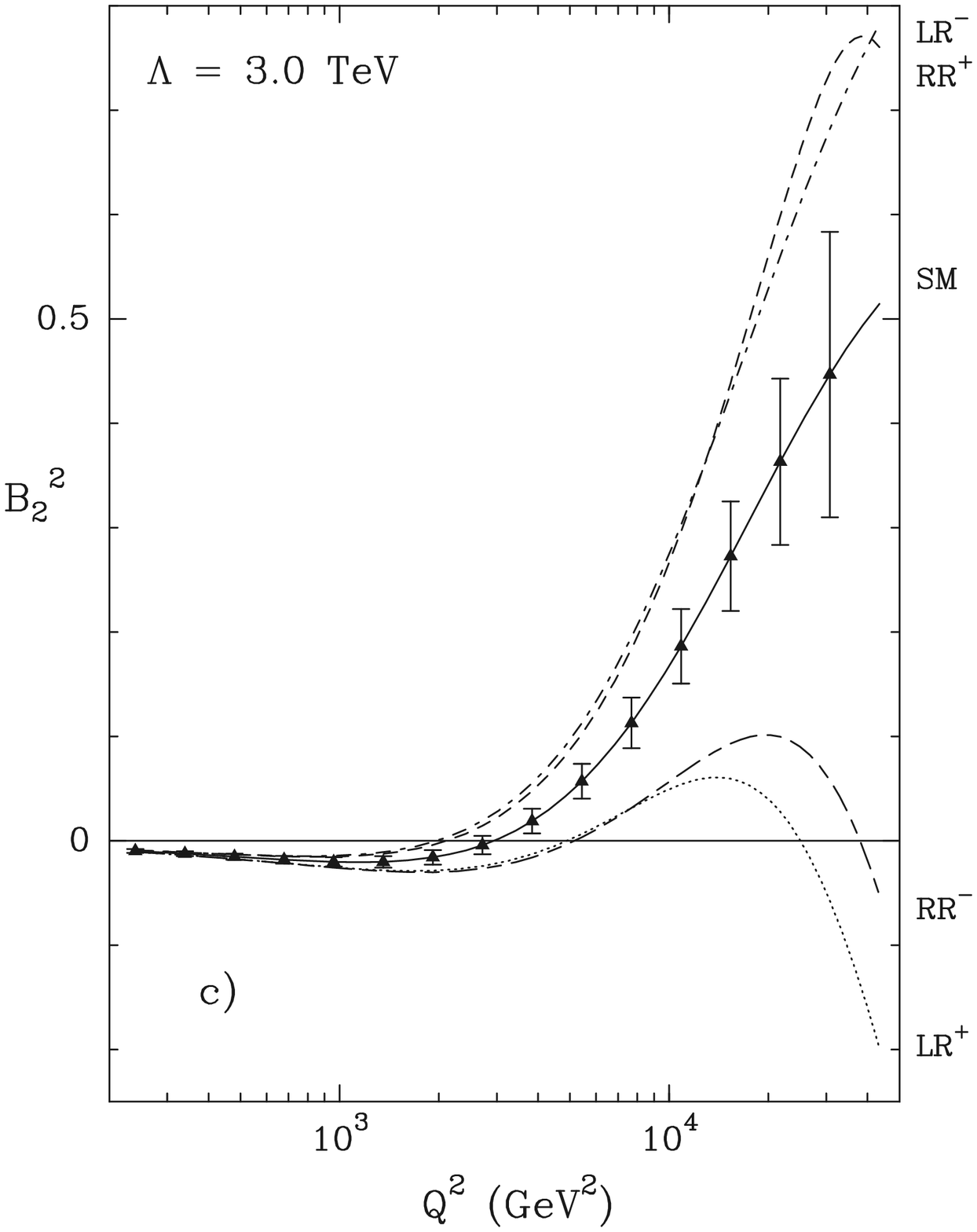},width=9truecm,height=13truecm}}\subfigureA{\psfig{file={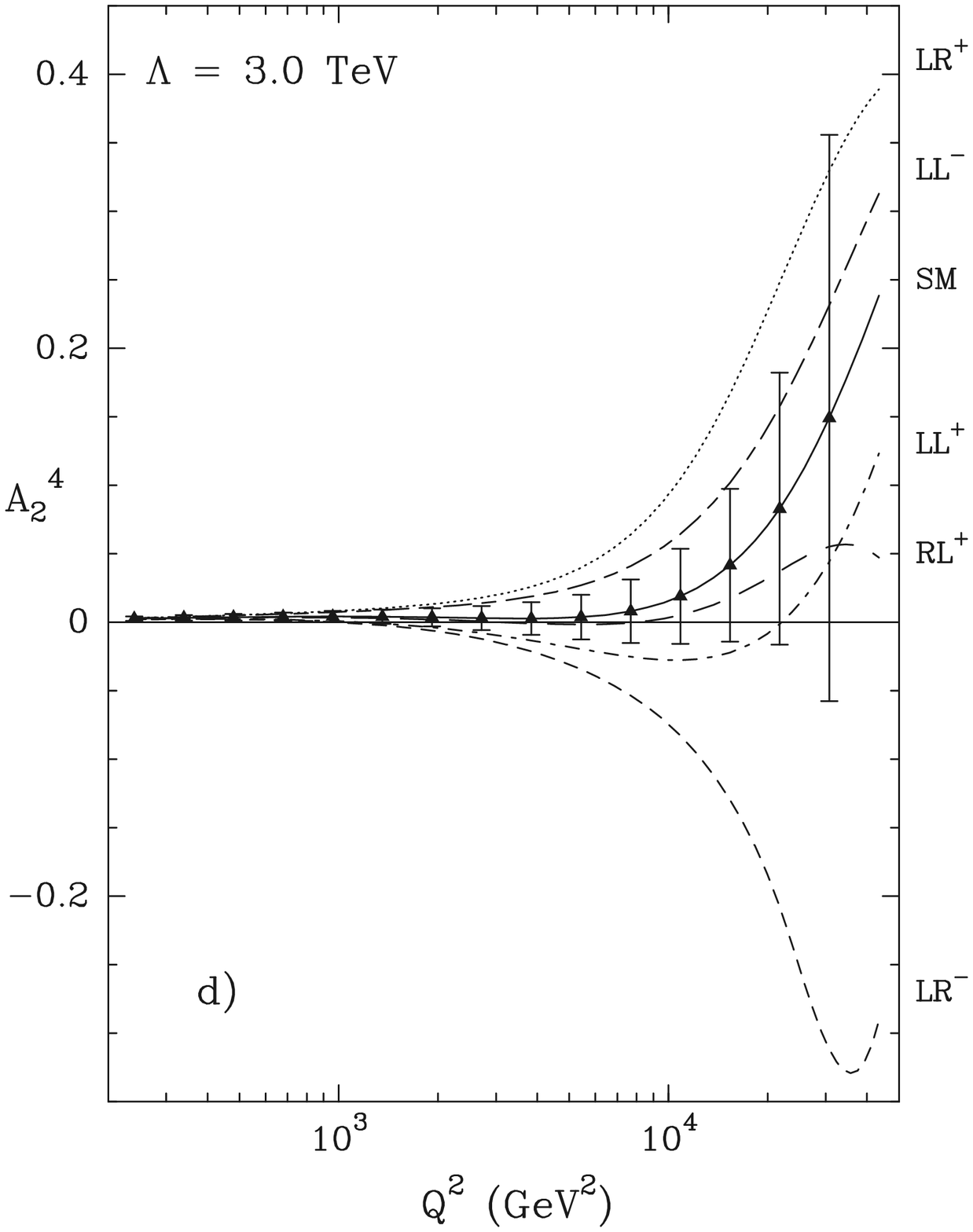},width=9truecm,height=13truecm}}}
\end{tabular} 
\caption{Spin asymmetries : a) $A_{LL}^{PV}(e^-)$, b) $A_{LL}^{PV}(e^+)$, c) $B_2^2$ and 
d) $A_2^4$ for the SM predictions (plain curves) and for the eight individual CI for the scale
$\Lambda = 3\, TeV$. The chiralities not mentionned are close to the SM.}
\end{figure}

\no The asymmetries $A_{LL}^{PV}(e^-)$ and $A_{LL}^{PV}(e^+)$ are represented on 
Fig.1.a-b, they are sensitive to the
chiralities ($LL^\pm,RR^\pm$) and ($LR^\pm,RL^\pm$), respectively. Now, the direction 
of the deviation from SM
expectations allows the distinction between two classes of chiralities. For instance,
a positive deviation for $A_{LL}^{PV}(e^-)$ pins down the class ($LL^+,RR^-$)
and, a negative one, the class ($LL^-,RR^+$). Similarly, an effect for 
$A_{LL}^{PV}(e^+)$ makes a distinction between ($LR^-,RL^+$) and ($LR^+,RL^-$).
We deduce that the measurement of
these two asymmetries would allow to separate the four classes :
\no ($LL^+,RR^-$), ($LL^-,RR^+$), ($LR^+,RL^-$) and ($LR^-,RL^+$).

We can go further in the identification of the chiral structure of the new interaction
by the use of additionnal asymmetries. For instance, $B_2^2$ is strongly sensitive
to the presence of the chiralities ($RR^\pm,LR^\pm$), see Fig.1.c. Again the direction of the 
deviation from SM distinguish ($RR^+,LR^-$) from ($RR^-,LR^+$). Since these
two classes are distinct from the four previous ones, we conclude that the measurements
of the three spin asymmetries $A_{LL}^{PV}(e^-)$, $A_{LL}^{PV}(e^+)$ and $B_2^2$
should give a clear identification of the chiral structure of the new interaction
{\it in this naive model}.

Now, it turns out that,
if the chiral structure of the new interaction is more complicated,
in general, measuring the three asymmetries mentionned above will be sufficient to 
identify the precise chiral structure. However, for some special cases, 
like for instance the $VV$ model \cite{Ruckl84} which conserves parity, 
some cancellations occur. Then
we need to measure some other spin asymmetries. It appears that the four PC
spin asymmetries, defined above, are particularly interesting, since they are roughly 
sensitive to one chirality only. For instance, the asymmetry $A_2^4$, which is mainly
sensitive to $LR^\pm$, is represented on Fig.1.d.
The problem of these PC asymmetries is that 
they are less sensitive to new physics than the PV one's (see Table 1). Then, if the
new interaction has a complicated structure, we can obtain some valuable informations
at a lower value of $\Lambda$ ($\sim 5$ TeV) only.

Finally, we can make some remarks on the one spin asymmetries defined
when only the lepton beams are polarized. The behaviour of these asymmetries
has been presented some years ago in \cite{Ruckl84}. It appears \cite{jmvp}
that these asymmetries are less sensitive to the presence of new physics
than the double spin asymmetries. The same behaviour has been noticed in the case of 
polarized $pp$ collisions at RHIC \cite{ptjmv}.
Moreover, we can't define as many asymmetries
as in the two spin case. For instance, we can not define the PC spin asymmetries.
Then, if the structure of the new interaction is complicated, we can loose
the opportunity to identify its chiral structure.


In conclusion, we can make a remark on the leptoquarks. Indeed, in order to be
phenomenologically acceptable at present or future experiments, we know from
pion decays and $(g-2)_\mu$ measurements that the leptoquarks must be of chiral nature
\cite{HR1}. Then we conclude that if a leptoquark is present and detectable at HERA,
it will certainly induce some deviations in the PV spin asymmetries
presented here.

\end{document}